\documentclass[10pt]{ourtemplate}
\usepackage{lipsum}
\usepackage{amsmath}
\usepackage[english]{babel}
\usepackage{float}
\usepackage{textcomp}
\usepackage{amsfonts}
\usepackage{amsthm}
\usepackage{graphicx}
\usepackage{dcolumn}
\usepackage{stackrel,amssymb}
\usepackage{color}
\usepackage[normalem]{ulem}
\usepackage[utf8]{inputenc}
\usepackage{hyperref}
\usepackage{graphicx}
\usepackage{makecell}
\usepackage[dvipsnames]{xcolor}
\usepackage{xcolor,colortbl}

\usepackage{float}
\usepackage{braket}
\usepackage{afterpage}

\usepackage{cite}
\usepackage{multicol}
\begin{document}

\justifying
\pagestyle{fancy}
\fancyhf{}
\cfoot{\thepage}
\renewcommand{\headrulewidth}{0pt}
\title{\bf Opportunities in Quantum Reservoir Computing and Extreme Learning Machines}

\begin{center}
\maketitle
\end{center}

\begin{flushleft}
\author{Pere Mujal, Rodrigo Mart{\'\i}nez-Pe{\~n}a, Johannes Nokkala, Jorge Garc{\'\i}a-Beni, Gian Luca Giorgi, Miguel C. Soriano, Roberta Zambrini*}

\begin{affiliations}

\normalsize
\noindent
Dr. P. Mujal, R. Mart{\'\i}nez-Pe{\~n}a, Dr. J. Nokkala, J. Garc{\'\i}a-Beni, Dr. G. L. Giorgi, Dr. M. C. Soriano, Prof. R. Zambrini\\
IFISC, Instituto de F\'{\i}sica Interdisciplinar y Sistemas Complejos (UIB-CSIC)\\
UIB Campus, E-07122 Palma de Mallorca, Spain\\
*Email Address: roberta@ifisc.uib-csic.es\\
Dr. J. Nokkala\\
Turku Centre for Quantum Physics, Department for Physics and Astronomy, \\University of Turku, FI-20014, Turun Yliopisto, Finland\\
\end{affiliations}

\end{flushleft}
\begin{multicols}{2}

\begin{abstract}
\noindent
\textbf{Quantum reservoir computing (QRC) and quantum extreme learning machines (QELM) are two emerging approaches that have demonstrated their potential both in classical and quantum machine learning tasks. They exploit the quantumness of physical systems combined with an easy training strategy, achieving an excellent performance.
The increasing interest in these unconventional computing approaches is fueled by the availability of diverse quantum platforms suitable for implementation and the theoretical progresses in the study of complex quantum systems.
In this review article, recent proposals and first experiments displaying a broad range of possibilities are reviewed when
quantum inputs, quantum physical substrates and quantum tasks are considered. 
The main focus is the performance of these approaches, on the advantages with respect to classical counterparts and opportunities.
}

\end{abstract}

\normalsize

%%%%%%%%%%%%%%%%%%%%%%%%%%%%%%%%%%%%%%%%%%%%%%%%%%%%%%%%%
\section{Introduction}

    In recent years, we are witnessing an explosion of unconventional computing methods and systems \cite{Adamatzky,jaeger2020exploring}.
    One of the driving motivations for these efforts on unconventional computing is to go beyond von Neumann architectures, physically co-locating processing and memory operations \cite{wright2013beyond,ielmini2018memory}.
    
    For the unconventional computing revolution to occur, computational models and computing substrates are to be considered as a whole.
    Neuromorphic, and more generally neuro-inspired, computing is one of such fields where the computational paradigm goes hand in hand with the design of the physical substrate, aiming at approaching the computational power of the human brain \cite{neuromorphic}.

    Machine learning, and in particular the field of artificial neural networks (NN), can similarly benefit from the progress in neuro-inspired computing devices \cite{benjamin2014neurogrid,furber2014spinnaker,indiveri2015memory}. The potential to build systems that are orders of magnitude more energy efficient than traditional ones is a major key motivation \cite{esser2016convolutional}. As computing approaches get closer to considerations on their physical substrates, the analog properties of physical systems come into focus \cite{indiveri2013integration}. Thanks to recent advances, artificial NN are envisioned to be run even on top of analog quantum computing devices \cite{Behrman2000,schuld2014quest,wright2020capacity}, with the possibility to exploit the advantages of superposition in quantum computing and the parallelism in neural computing. The implications of combining machine learning and quantum physics indeed represents a major avenue for research in the coming years~\cite{biamonte2017quantum,Dunjko_2018,RevModPhys.91.045002,doi:10.1063/5.0020014,PhysToday72,Ciliberto2018}.
    
\begin{figure}[H]
    \centering
    \includegraphics[width=0.9\columnwidth]{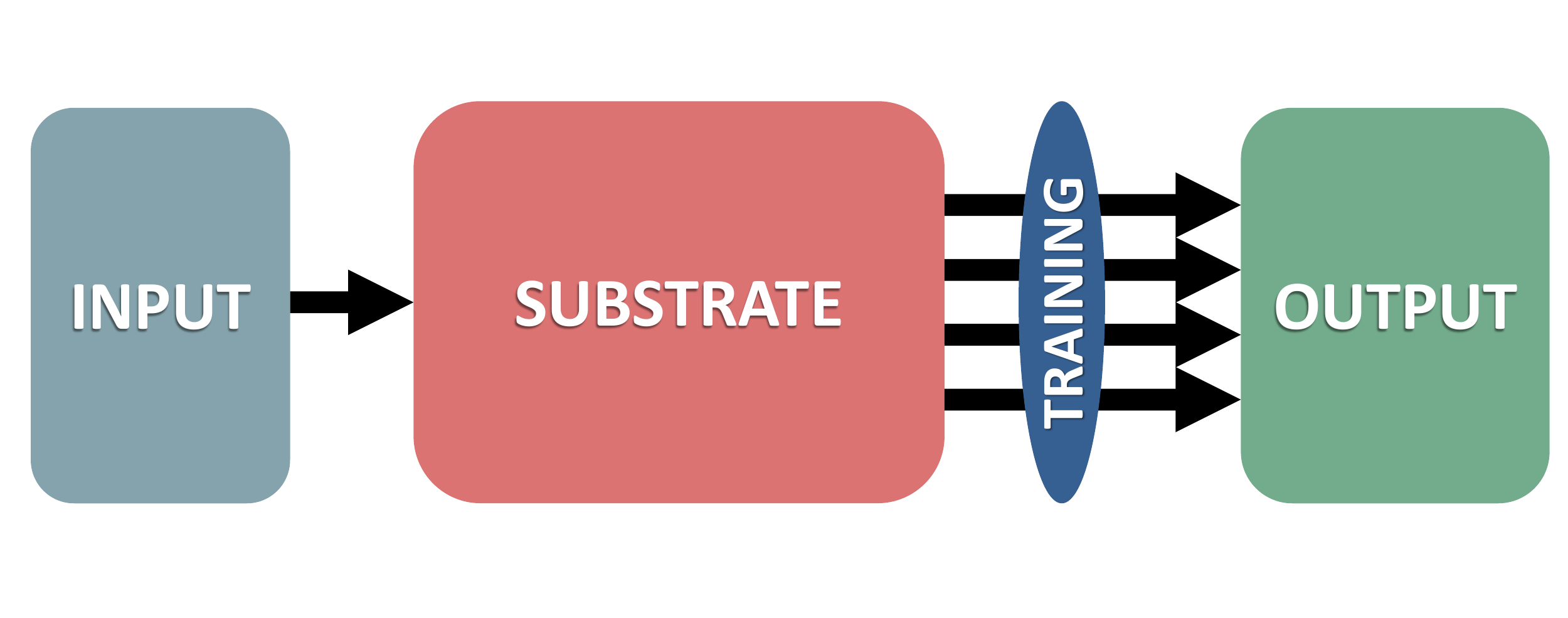}
    \caption{Schematic representation of the basic ingredients for RC and ELM. The information from the input is fed into the substrate, which acts as a hidden layer or reservoir. The response of the substrate, %i.e. 
    through a selection of observables, is then used to produce the desired output after optimization of the output connections by training.}
    \label{fig1}
\end{figure}

    In this review article, we concentrate on the potential of quantum devices for reservoir computing (RC) and extreme learning machines (ELM). RC and ELM are machine learning paradigms that exploit the natural dynamics of input-driven randomly connected NN for information processing \cite{lukovsevivcius2009reservoir,huang2011extreme}. The main advantage of the RC and ELM concepts in the context of artificial NN is their minimal requirements for learning (usually referred as training in the machine learning literature) \cite{lukovsevivcius2012practical,butcher2013reservoir}. Figure~\ref{fig1} illustrates the three layers typical of RC and ELM, namely an input layer, a hidden layer or substrate, and an output layer. In RC and ELM, one only needs to adjust the weights of the output connections via, for example, linear regression (see Figure~\ref{fig1}), while the rest of the connections can be initialized with random weights and are not optimized. Despite the simplicity in the training, these methods have been successful in numerous practical applications \cite{lukovsevivcius2012reservoir,huang2015trends,beque2017extreme,antonik2019human,PhysRevX.10.041037}.
    Although RC and ELM are both random mapping architectures \cite{butcher2013reservoir}, a main difference resides in the fact that RC exploits the natural dynamics of the substrate as an internal memory of past
    input information while ELM does not. Both RC and ELM are amenable to dedicated hardware implementations in, for example, digital electronics \cite{haynes2015reservoir,frances2016hardware} or nonlinear analog systems \cite{nakajima2020physical,ortin2015unified}. The quantum counterparts of RC and ELM will be referred to in the following as QRC and QELM, respectively.
    As common in the literature on these topics,
    we consider as quantum RC and ELM if they are based on quantum substrates and this will be the main focus of this article. We note that this is a more restrictive definition with respect to what is usually
\noindent 
considered as quantum in machine learning (where either data or computing device can be non-classical).

Quantum systems exhibit a large number of degrees of freedom that can be exploited for QRC and QELM. Since the pioneering proposal of networks of quantum spins as reservoir substrates \cite{PhysRevApplied.8.024030}, there have been a variety of works exploring the possibilities that quantum mechanics can offer to this research area. In particular, the spin-based implementation is the most analyzed platform for QRC at
the moment \cite{PhysRevApplied.11.034021,Kutvonen2020,higherorderqrc,Chen2019,9029180,PhysRevApplied.14.024065,martinez2020information,nisqstockforecast,martinez2021},
with the continuous-variable systems becoming another promising option~\cite{nokkala2020gaussian,kalfus2021neuromorphic,IEEEpaper}.
In the context of QELM, both fermionic and bosonic setups have been proposed for instance in entanglement detection \cite{Ghosh2019} or light field phase estimation \cite{qrcsnonlinearoscill}. Inspired by these neuromorphic approaches other quantum tasks have been reported such as  quantum states
preparation \cite{PhysRevLett.123.260404,KRISNANDA2021141} or  reconstruction \cite{9153954}. Proof-of-principle experiments are also ongoing  \cite{PhysRevApplied.14.024065,negoro2018machine}. In particular, QELM has been reported in a NMR experiment \cite{negoro2018machine}, while QRC has been realized on the quantum computation platform of IBM  \cite{Chen2019,9029180, PhysRevApplied.14.024065}.

Exploring extended quantum systems as substrates for machine learning in QRC and QELM
represents a timely and potentially disruptive opportunity. 
First of all, given a substrate of N nodes, the exponential size of the Hilbert space allows in principle for a substantially enlarged output when compared to classical ones \cite{PhysRevApplied.8.024030}. The extent of this advantage when accessing a large number of output degrees of freedom has been quantified for instance in QRC with spin \cite{martinez2020information} as well as Gaussian networks \cite{nokkala2020gaussian}. While decoherence is a strong limiting factor in gate-based quantum computing, in principle  QRC and QELM are well suited for NISQ \cite{qrsmlnisq}. Furthermore quantum substrates can interact with quantum inputs and are naturally suited for quantum tasks, providing new avenues for edge computing in quantum systems with increasing complexity. Beyond the opportunities in the context of quantum technologies for computation with a quantum advantage,  QRC and QELM offer a new and original perspective to characterize quantum physical systems in terms of their operation as substrates \cite{martinez2021}.

In the context of quantum machine learning, four approaches  are often identified referring to the different combinations of classical (C) or quantum (Q) data as well as to the classical (C) or quantum (Q) computing devices \cite{Aimeur2013}.
This classification in either CC, CQ, QC or QQ (data/device) is also useful to frame the recent advances in QRC and QELM.
Here we will further label the classical or quantum nature of the computational task. This input/substrate/task classification will guide the discussion of existing approaches in this review article. 

The rest of the article is organized as follows. In Section~\ref{mainsec}, we give a brief description of the mathematical formalism for the full classical case and we comment on how to extend it to the quantum case. In Section~\ref{iputencoding} and ~\ref{tasks}, we concentrate on the different input forms and their treatment in order to do a given computational task, respectively. In Section~\ref{quantumsubstrates}, we analyze the substrates that have been proposed and their requirements. The performance of QRC and QELM in some exemplifying works is underlined in Section~\ref{examples} and \ref{performance}. In Section~\ref{challenges}, we point out the current challenges in both experimental and theoretical contexts. We end the article by highlighting potential opportunities for this research field.
% 
%%%%%%%%%%%%%%%%%%%%%%%%%%%%%%%%%%%%%%%%
\section{Quantum Resources for Unconventional Computing}
\label{mainsec}

\subsection{Reservoir Computing and Extreme Learning Machines}
\label{RCandELM}

In this section, we provide an overall picture of the formalism behind RC and ELM before moving on to their quantum counterparts in the following sections.
We will use this general framework to assess previous works in QRC and QELM and provide an umbrella for the classification of future works.

All possible situations regarding the classical (C) or quantum (Q) character of the input, the substrate and the task, respectively are summarized in Table~\ref{table1}. Traditional RC and ELM fall into the first column of the table, whereas QRC and QELM correspond to the second one.
For instance, the first proposal of QRC~\cite{PhysRevApplied.8.024030} belongs to the CQC class: \textit{a C input is fed into a Q substrate to do a C task}. We will elaborate on the content of Table~\ref{table1} along the following sections.
The overview provided by the classification in the table reflects the richness of possibilities and therefore of opportunities in this research field.

%\threesubsection
\subsubsection{Classical Reservoir Computing}\label{crcomp}

In general, RC deals with (physical or artificial) dynamical systems that can map a given input to their state space \cite{lukovsevivcius2009reservoir,brunner2019photonic}. Here we briefly review the  RC definition in the broadly studied CCC scenario. If the states are vectors of $N$ real numbers, $\mathbf{x}_k$, and inputs sequential real scalars  (for simplicity), $\{s_k\}$, then the most general form of such a map would be \cite{konkoli2017reservoir}
\begin{equation}
\label{eqmapRC}
\mathbf{x}_k=\mathbf{f}(s_k,\mathbf{x}_{k-1}),  
\end{equation}
where $\mathbf{f}$ is a fixed function determined by, e.g., the system itself and how the input is injected, and, in general, can
%%%%%%%%%%%%%%%%%%%%%%%%%
\begin{table}[H]
\caption{All possible combinations of input, substrate and task being classical (C) or quantum (Q). The sub-index $k$ labels each time step or instance.
The sub-index $i$ is associated to the internal degrees of freedom of the substrate.
For the classical input, $\{s_k\}$ is a data sequence, e.g. a string of real numbers. In the quantum case, $\rho_k^{\textnormal{in}}$ is a density matrix representing an input state. The state of the substrate at a given instant is defined by $\mathbf{x}_k$ in the classical regime and by the density matrix $\rho_R$ in the quantum one.
For the training process with a classical substrate, a selection of the substrate variables are used, $\mathbf{x}_k^{\textnormal{out}}$. With a quantum substrate, the readout for the training is obtained after a set of measurements, $\{O_i^{\textnormal{out}}\}$. We distinguish between classical tasks, $T_{\textnormal{C}}$, and quantum tasks, $T_{\textnormal{Q}}$.}
\setlength\arrayrulewidth{1.5pt}\arrayrulecolor{white}
\begin{center}
\resizebox{\columnwidth}{!}{
\begin{tabular}{|c|c|c|}
%\arrayrulecolor{black}
\cline{2-3}
\multicolumn{1}{c|}{\makecell{\, \\ \, \\ \,}}&
\multicolumn{1}{c|}{\cellcolor{ForestGreen!30} \makecell{{\bf Classical Substrate} \\ $\mathbf{x}_k$}} & \cellcolor{blue!25} \makecell{\,{\bf Quantum Substrate}\, \\ $\rho_{\textnormal{S}}$} \\
%\arrayrulecolor{black}
\hline
\cellcolor{ForestGreen!30} \makecell{{\bf \,Classical\,} \\ {\bf Input}}  & \cellcolor{ForestGreen!15}\makecell{{\bf CCC} \\ \,\,\,\,\,\,\,\,$\{s_k\},\mathbf{x}_k^{\textnormal{out}}\rightarrow T_{\textnormal{C}}$\,\,\,\,\,\,\,\,\, \\
\cite{TANAKA2019100,butcher2013reservoir,konkoli2017reservoir,jaeger2001echo,maass2002real,grigoryeva2018echo,huang2006extreme,huang2011extreme,marcucci2020theory}} & \cellcolor{ForestGreen!15} \makecell{{\bf CQC} \\ \,\,\,\,\,\,\,$\{s_k\},\{O_i^{\textnormal{out}}\}\rightarrow T_{\textnormal{C}}$\,\,\,\,\, \\ \cite{PhysRevApplied.8.024030,PhysRevApplied.11.034021,Kutvonen2020,PhysRevApplied.14.024065,Chen2019,9029180,qrcsnonlinearoscill,kalfus2021neuromorphic,higherorderqrc,qrsmlnisq,nisqstockforecast,martinez2020information,nokkala2020gaussian} } \\
%\cline{2-3}
\cellcolor{ForestGreen!30} $s_k$  & \cellcolor{blue!15} \makecell{{\bf CCQ}\\ \,\,\,\,\,\,\,\,$\{s_k\},\mathbf{x}_k^{\textnormal{out}}\rightarrow T_{\textnormal{Q}}${\bf \,\,\,\,}\,\,} & \cellcolor{blue!15}\makecell{{\bf CQQ} \\ \,\,\,\,\,\,\,\,$\{s_k\},\{O_i^{\textnormal{out}}\}\rightarrow T_{\textnormal{Q}}$\,{\bf\,\,\,\,} \\ \cite{PhysRevLett.123.260404,KRISNANDA2021141}} \\
\hline
\cellcolor{blue!25} \makecell{{\bf Quantum} \\ {\bf Input}}  & \cellcolor{ForestGreen!15} \makecell{{\bf QCC} \\ \,\,\,\,\,\,\,$\{\rho_k^{\textnormal{in}}\},\mathbf{x}_k^{\textnormal{out}}\rightarrow T_{\textnormal{C}}$\,\,\,\,\,\,} & \cellcolor{ForestGreen!15} \makecell{{\bf QQC} \\ \,\,\,\,\,$\{\rho_k^{\textnormal{in}}\},\{O_i^{\textnormal{out}}\}\rightarrow T_{\textnormal{C}}$\,\,\,\,\,} \\
%\cline{2-3}
\cellcolor{blue!25} $\rho_k^{\textnormal{in}}$  & \cellcolor{blue!15} \makecell{{\bf QCQ}\\ \,\,\,\,\,\,\,$\{\rho_k^{\textnormal{in}}\},\mathbf{x}_k^{\textnormal{out}}\rightarrow T_{\textnormal{Q}}${\bf\,\,\,\,\,} \\ \cite{rcqsmeasurement}} & \cellcolor{blue!15} \makecell{{\bf QQQ} \\ \,\,\,\,\,$\{\rho_k^{\textnormal{in}}\},\{O_i^{\textnormal{out}}\}\rightarrow T_{\textnormal{Q}}${\bf \,\,\,\,}\\ \cite{universalqrc,9153954,Ghosh2019,Kawai2020}} \\ \hline
\end{tabular}
}
\end{center}
\label{table1}
\end{table}
%%%%%%%%%%%%%%%%%
\noindent
have linear and nonlinear contributions \cite{jaeger2001echo,lukovsevivcius2009reservoir,paudel2020classification}.
Notice that the state of the substrate at a given time, $\mathbf{x}_k$, depends also on past inputs. This is seen recurrently iterating Eq.~\ref{eqmapRC}, e.g. for three steps $\mathbf{x}_{k}=\mathbf{f}\left(s_{k},\mathbf{f}(s_{k-1},\mathbf{f}(s_{k-2},\mathbf{x}_{k-3}))\right).$
As an illustration, the input might be a temporal signal such as speech, and the corresponding task might be to interpret oral conversations into text in real time. As is typical for RC, in this example each desired output (words) depends on multiple past inputs (recent sounds), i.e. the input history, $s_{i\leq k}$, so the state of the system is used as memory.
Additionally, the map $\mathbf{f}$ should be contracting such that $\mathbf{x}_k$ to a good approximation depends only on recent dynamics and  input history \cite{jaeger2001echo,maass2002real,grigoryeva2018echo}; this ensures both convergence (so that input sequences that coincide during some recent time lead to the same outputs, a property also known as fading memory) and independence of distant past such as the initial state of the system (echo state property). Finally, the map should also guarantee separability \cite{konkoli2017reservoir}, which implies that the reservoir computer separates any pair of different inputs into different outputs.

RC builds on this simple approach to solve machine learning tasks: the bulk of the processing is offloaded to a fixed complex system and desired input-output maps are achieved by only adjusting how its state is post-processed---this is how the output of such systems is trained and a common feature of random mapping methods \cite{butcher2013reservoir}.
The output sequence, $\{y_k\}$, is obtained from some of the state variables of the substrate, $\mathbf{x}_k^{\textnormal{out}}$, 
\begin{equation}
y_k=h(\mathbf{x}_k^{\textnormal{out}}),
\end{equation}
where $\mathbf{x}_k^{\textnormal{out}}$ is a vector of $L\leq N$ states, i.e. all state variables may not be accessible for the output, and the function $h$ contains free parameters that are optimized during the training process. The supervised learning is done by minimizing a cost function between the true target output, $\bar{y}_k$, and the predicted one, $y_k$. It is worth mentioning that the generalization of this framework to a vectorial input, $\{\mathbf{s}_k\}$, or output, $\{\mathbf{y}_k\}$, is straightforward~\cite{PhysRevLett.120.024102,PhysRevX.10.041037}.

The overall objective of RC is to provide a high-dimensional mapping of the input $s_k$ to $\mathbf{x}_k$, and ultimately approximate the desired output $y_k$. To this end, the functions $\mathbf{f}$ and $h$ provide memory and usually define a nonlinear transformation from input to output.
Memory and non-linearity resources play a key and distinct role in different tasks.
A useful quantifier of the memory of a given system, used also in works considering quantum reservoirs as will be seen, is called information processing capacity (IPC)~\cite{dambre2012information}.
The IPC facilitates the analysis of what kind of linear and nonlinear memory functions a given system can approximate. Further details are provided in Appendix~\ref{app_IPC}.
\,\\

\subsubsection{Classical Extreme Learning Machines}

The main difference between ELM and RC is that in an ELM the state of the substrate, $\mathbf{x}_l$, is uniquely defined by only its corresponding input, $s_l$. Therefore, the map between the two is given by~\cite{huang2006extreme,huang2011extreme,marcucci2020theory}
\begin{equation}
\mathbf{x}_l=\mathbf{f}(s_l),
\end{equation}
where $l$ denotes different instances of the input. In ELM, temporal dependencies between different input instances are not captured and are irrelevant, so it specializes in a different kind of tasks. The set of inputs, $\{s_l\}$, is not given in any relevant order. For instance, in a classification problem, in which we want to know if there is a car in a picture or not, the presence of a car in the previous picture is not related to the current one.
It should be stressed that a RC can be used as an ELM by, e.g., simply re-initializing it between inputs, and an ELM can be used as a RC provided that $\mathbf{f}$ is contracting as described above and the state is allowed to retain memory of past inputs. Indeed, situations can be envisioned where the same physical system can be used for both \cite{ortin2015unified}.

In the following sections, we analyze and discuss the extension of the two approaches to the quantum domain, i.e. QRC and QELM, where the substrate is a quantum system. We start discussing how to encode the input information, which needs to trigger the substrate properly. We then discuss recent  proposals of quantum tasks as well as some requirements of a physical system to be used as a substrate. Once the main elements of QRC and QELM have been introduced, two relative examples of tasks --namely timer and classification-- will be shown  in Section~\ref{examples}. These tasks are common in the classical framework and we will show them for a quantum reservoir of spins and of harmonic oscillators, respectively.

%%%%%%%%%%%%%%%%%
\subsection{Input Encoding} 
\label{iputencoding}

Once we have a computational problem to be tackled with the use of a classical or quantum substrate as a processing unit, the first issue to be addressed is how to introduce the input into the system and different strategies have been devised. One possibility, as we explain in more detail below, is by directly accessing to all or a part of the substrate. Alternatively, an external system can serve as an ancilla and be used to encode the input, which is then coupled to the substrate. In both cases we can distinguish between classical inputs and quantum ones, as consistently classified in Table~\ref{table1}, indicating them through a scalar or a quantum state.

When the input is classical, the information is usually given as a series of real numbers, $\{s_k\}$, with $k=0,1,\,...\,$, as explained in Section \ref{RCandELM}. For QRC, each instance of the input is fed into the substrate at consecutive time steps, $\Delta t$, so that time is given by $t=k\Delta t$, whereas for a QELM task, the time does not play a relevant role and $k$ represents each instance.

In the first proposal of QRC~\cite{PhysRevApplied.8.024030}, the designed algorithm is suited for processing classical temporal information, e.g. given a sequential input $\{s_k\}$ the goal is to reproduce a classical target output $\{y_k\}$.
In order to input the classical sequence into the quantum substrate (of interacting qubits) the input is first encoded in the quantum state through one (or more) of the qubits. The state of this selected qubit is recurrently prepared in a different superposition of the basis states $\ket{0}$ and $\ket{1}$. The value of the classical input, $s_k \in [0,1]$ for a normalized continuous variable, or $s_k \in \{0,1\}$ for a binary one, fixes the components, i.e. $\ket{\psi_k}=\sqrt{1-s_k}\ket{0}+\sqrt{s_k}\ket{1}$.
Subsequent works concerning time-series predictions have employed a similar input encoding procedure~\cite{Kutvonen2020,martinez2020information} in pure states.
An alternative way to encode the classical input is used in~\cite{PhysRevApplied.11.034021,PhysRevApplied.14.024065,Chen2019,9029180,higherorderqrc,qrsmlnisq}, where the input qubit is consecutively initialized in the general mixed state $\rho_k=(1-s_k)\ket{0}\bra{0}+s_k\ket{1}\bra{1}$. The state is pure when $s_k=0$ or $s_k=1$, and is completely mixed for $s_k=1/2$. 
In continuous variable systems, the input can be encoded, for instance, in coherent states amplitude, in squeezing strength or phase, in thermal excitation. These encodings have been analyzed in the case of single-mode Gaussian states in harmonic networks in Ref. \cite{nokkala2020gaussian}, where it was shown that the encoding choice can display a significant effect on the QRC performance (see Section~\ref{performance}).

Another less explored possibility to fed a classical input into a quantum substrate is by driving the substrate with an external field, such as a light beam modeled by a time-dependent parameter in the Hamiltonian that describes the reservoir.
As an example, the input of the anharmonic oscillator processing system in Refs.  \cite{qrcsnonlinearoscill,kalfus2021neuromorphic} is the phase of a monochromatic light beam driving of the dynamics. Each phase instance is learned through measurements at different times, while the system is reset between inputs, representing an example of QELM with time multiplexing. 

Some approaches inspired by QRC and QELM have also addressed problems where the input is quantum, being an ancilla mode in a quantum state.
The dynamical evolution of the density matrix of the coupled quantum substrate and the quantum input modes enables different tasks~\cite{9153954,Ghosh2019}.
For instance, quantum input states can be classified when encoded into an ancilla interacting with a fermionic network  ~\cite{Ghosh2019}.
Further, the input quantum state, either in finite dimension or in continuous variable, can be reconstructed \cite{9153954}.
Recently, a quantum input has been also considered in classical RC. For instance, in continuously monitored superconducting qubits~\cite{PhysRevX.10.011006,rcqsmeasurement}, the information about the quantum state is carried by a microwave signal resulting from the measurement process.

%%%%%%%%%%%%%%%%%
\subsection{Computational Tasks}
\label{tasks}

In this section, we will describe recently proposed computational tasks that are suitable for QRC and QELM. In the first place, it is worth introducing a criterion to establish whether a given task is classical or quantum.
We consider a task  to be quantum when it exploits the quantum information content of input and/or output states. Examples are the identification of quantum correlations or quantum state preparation. This definition is adopted in Table~\ref{table1}.
We notice that, classical, $T_{\textnormal{C}}$, and quantum, $T_{\textnormal{Q}}$, tasks can be performed with both classical and quantum substrates. 

Starting with classical tasks, there are several recent studies about the performance of quantum reservoirs as substrates for (classical) time-series processing~\cite{PhysRevApplied.8.024030,PhysRevApplied.11.034021,Kutvonen2020,PhysRevApplied.14.024065,Chen2019,9029180,qrcsnonlinearoscill,kalfus2021neuromorphic,higherorderqrc,qrsmlnisq,nisqstockforecast,martinez2020information,nokkala2020gaussian,martinez2021}, so they are in the CQC class. 
These include benchmark tasks commonly considered in classical RC such as the timer task (see Section~\ref{sec:QRCexample}), realizing nonlinear functions of past inputs such as the nonlinear autoregressive moving average (NARMA) \cite{atiya2000new}, and chaotic time series prediction based on, e.g. the Mackey-Glass system \cite{jaeger2004harnessing}. As for QELM, there are at present no attempts to apply quantum reservoirs to classical tasks such as, for instance,  classification of images ~\cite{huang2006extreme}. Further examples of classification tasks have been reported in other NN settings \cite{classific_QNN}  
as well as in recently reported quantum classifiers
\cite{havlivcek2019supervised,quantum_classifier_2}. 
This CQC framework corresponds to the extension of the full classical approach~\cite{TANAKA2019100,butcher2013reservoir,konkoli2017reservoir,jaeger2001echo,maass2002real,grigoryeva2018echo,huang2006extreme,huang2011extreme,marcucci2020theory} by replacing the classical substrate with a quantum one. The main goal is the realization of similar classical tasks with an improved performance (Section~\ref{performance}).

There are also recent proposals of quantum tasks that can be considered as QELM or inspired by it.
In Ref.~\cite{Ghosh2019} the quantum task consists in the detection of entanglement and the computation of entan\-glement-related quantities, which are typically difficult to extract from experimental setups. The setup consists of a quantum input ancilla coupled to a quantum network.
Similarly, also quantum state tomography based on a quantum reservoir has been reported, another example of a QQQ protocol in a framework similar to QELM. Conventional quantum tomography schemes are challenging but,
with the quantum reservoir approach in~\cite{9153954}, the unknown density matrix of the input quantum state can be reconstructed after a single measurement on  local observables of the reservoir nodes without the need of any correlation detection.
The characterization of quantum state is indeed an important quantum task, also explored in other platforms, like for instance the quantum neuron proposed in Ref. \cite{kristensen2019artificial}.

Beyond detection schemes, another example of a $T_{\textnormal{Q}}$ is the preparation of desired quantum states~\cite{PhysRevLett.123.260404,KRISNANDA2021141} in ancilla systems interacting with different quantum substrates. In~\cite{PhysRevLett.123.260404}, for instance anti-bunched and cat states have been reported while Ref.~\cite{KRISNANDA2021141} addresses maximally entangled states, NOON, W, cluster, and discorded states. 
The main difference with QRC and QELM frameworks is that in spite of the presence of an input signal in the system, this is just driving it in a suitable operation regime but it is not an information input: the classical input does not encode which is the target quantum state, being this independently imposed through the proper choice of a cost function. 
As no quantum resource is needed in the input, in Table~\ref{table1}, these examples are listed in the CQQ category.
As a note, a QQC protocol could be devised  as the inverse process to the previous one. For instance, given a generic (quantum) state of an ancilla system as input, the task could be to find a classical control parameter in the substrate Hamiltonian preserving that quantum state.

A QELM approach has also been proposed for quantum chemistry calculations in ~\cite{Kawai2020}, being an example of the QQQ class. The first two excited molecular energies and transition dipole moments between the ground states and the corresponding excited states are predicted from the ground-state wavefunction of the molecule, which is given as an input to a quantum reservoir formed by a random transverse-field Ising model, acting as an entangler circuit of qubits.

As shown in \cite{universalqrc}, a quantum substrate  can also provide a general framework for quantum computing, achieving a universal set of quantum gates
and also non-unitary operations, of interest to simulate open quantum systems. Although this is a prominent quantum task (QQQ)  inspired by extreme learning techniques, the  model of Ref.~\cite{universalqrc} escapes  the basic three-layer scheme of Figure~\ref{fig1} with random or arbitrarily chosen input mappings. Indeed, in \cite{universalqrc} the weight optimization
involves both the input-reservoir coupling and the reservoir-output links, given that the physical input nodes coincide with the output ones. 
While this kind of mirror architecture is specific of the cited work, we notice  %argue 
that any task whose target is represented by a time series of density matrices would require some tuning of (part of) the Hamiltonian parameters.

QRC has also been considered in the context of quantum state measurement,
as a strategy for processing information coming from continuously monitored superconducting qubits coupled to a Kerr network acting as a reservoir~\cite{rcqsmeasurement}.
The substrate of Josephson parametric oscillators is operating into a classical regime and then classified as QCQ in Table 1.  We also mention a recent experiment
%%%%%%%%%%%%%%%%%%%%%%%%%%%%%%%%%%%%%%%%%%%%%%%%%%%%%%%%%
%
\begin{figure}[H]
\centering
\includegraphics[width=0.9\columnwidth]{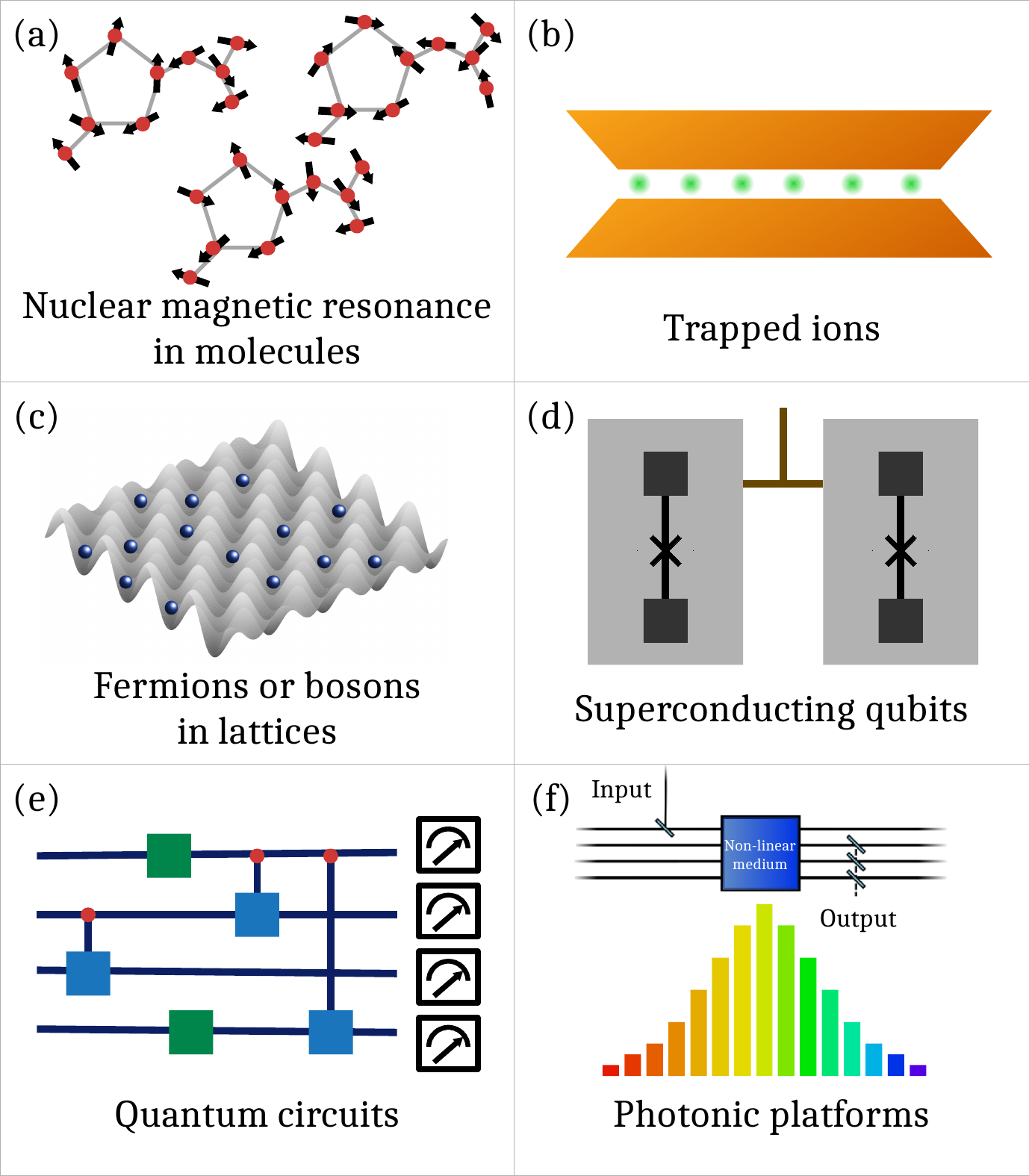}
\caption{(a)-(f) Various examples of platforms to be used as quantum substrates for information processing. Two suitable candidates to implement spin-network models~\cite{PhysRevApplied.8.024030,PhysRevApplied.11.034021,Kutvonen2020,Chen2019,higherorderqrc,qrsmlnisq,martinez2020information} are
(a) nuclear magnetic resonance in molecules~\cite{negoro2018machine}, and (b) trapped ions \cite{pino2020demonstration}.
(c) State-of-the-art ultracold atomic setups in optical lattices with bosonic and fermionic species are well-suited for Bose-Hubbard and Fermi-Hubbard discrete models~\cite{universalqrc,9153954,Ghosh2019,PhysRevLett.123.260404,KRISNANDA2021141}, respectively.
Additionally, other physical implementations are possible, e.g. in arrays of quantum dots in semiconductor devices, and in (d) coupled superconducting qubits~\cite{havlivcek2019supervised}.
(e) Quantum circuits~\cite{9029180,killoran2019continuous} have been used as a substrate in the IBM platform~\cite{PhysRevApplied.14.024065,nisqstockforecast}.
(f) Continuous-variable models~\cite{nokkala2020gaussian,qrcsnonlinearoscill,killoran2019continuous} could be engineered in photonic experimental setups as well~\cite{Kollar2019,Nokkala_2018njp,Cai2017}, i.e. by coupling different frequency modes in non-linear media.
}
\label{fig2}
\end{figure}    
%
%%%%%%%%%%%%%%%%%%%%%%%%%%%%%%%%%%%%%%%%%%%%%%%%%%%%%%%%%
\noindent
with a single qubit coupled to a superconducting cavity whose output is fed into a recurrent NN (implemented in software)~\cite{PhysRevX.10.011006}. The NN was able to predict measurement outcome probabilities
as well as to estimate the initial state of the qubit and parameters such as Rabi frequency, dephasing rate and measurement rate. 

\subsection{Quantum Substrates for Information Processing}
\label{quantumsubstrates}

In QRC and QELM, quantum dynamical systems amenable to input driving and output extraction 
are exploited to solve the tasks at hand. The exponential growth of Hilbert space dimension obtained by increasing the number of quantum elements endows them with a large state space that plays a determinant role in the performance \cite{PhysRevApplied.8.024030}. Such quantum substrates require several ingredients to solve computational tasks.

The first one is the need of a rich spatial and/or temporal dynamical behavior, which is frequently achieved introducing disorder in the interactions between different system components \cite{nakajima2020physical}. In fact, symmetries in the quantum substrate can cause the appearance of local conserved quantities that would reduce the number of exploitable degrees of freedom, which are responsible for a richer dynamics \cite{martinez2021}.
The second one has to do with how the state of the substrate depends on the input. As in the classical case, QRC requires contractiveness. Contraction of the dynamics, i.e., the convergence of two different initial states after repeated input injections, guarantees that the system presents a fading memory and allows the proper learning of temporal functions \cite{grigoryeva2018echo}. In this manner, initial conditions of the quantum system are erased and the state of the system only depends on the input history. At variance, in QELM it is usually assumed that the substrate is initialized into some fixed state before an input is processed. In this case, the state depends on the most recent input only. Finally, the state of the system should be nonlinear in input to achieve nontrivial information processing with a simple---and consequently easy to train---readout \cite{lukovsevivcius2009reservoir}.

With these requirements in mind, different systems and models have been proposed that could be implemented in several platforms. 
Figure~\ref{fig2} illustrates a selection of the physical platforms that are considered for QRC and QELM. Experimental implementations of such quantum substrates for QRC and QELM can be modeled as few-body systems.

\subsubsection{Spin and Other Discrete Variable Reservoirs}

The first proposal for QRC was based on quantum spin networks \cite{PhysRevApplied.8.024030}. 
The size of the Hilbert space of these spin-based systems grows as $2^N$, where $N$ is the number of spins, but the total number of degrees of freedom amounts to $4^N-1$. Then, NISQ technologies with tens of spins ensure a large state space. 
In this first work \cite{PhysRevApplied.8.024030}, the rich and complex dynamics of the transverse-field Ising model is evaluated in different numerical experiments and 
a competitive performance is obtained by just using the single spin projections in the $z$ direction $\braket{\sigma^z_j}$, while more and different observables have been addressed in \cite{martinez2020information}. Such performance partly comes from the exploitation of a technique called temporal multiplexing, where a number of $V$ evenly spaced snapshots of the dynamics are taken between two input injections, thus increasing the number of computational nodes by a factor of $V$ \cite{PhysRevApplied.8.024030, martinez2020information}.
The spin-based approach has since been further refined. In \cite{PhysRevApplied.11.034021}, spatial multiplexing is introduced, where multiple networks receive the same input and output is gathered from all of them
in a single output layer. In \cite{higherorderqrc}, spatial multiplexing is combined with classical connections between the networks to enable multidimensional inputs.
Nuclear magnetic resonance in molecules, as depicted in Figure~\ref{fig2}(a), and trapped ions, as depicted in Figure~\ref{fig2}(b), are potential platforms for spin-based QRC and QELM systems. 

Most works on spin-based systems for QRC mainly explore their applicability and computational performance from a numerical perspective, but efforts in the analytical direction have also been made. One of the main theoretical problems in the field of machine learning is the universal approximation property: the possibility to find examples of a class of systems that can approximate elements of a given class of functions with arbitrary precision.
In Ref. \cite{Chen2019}, focusing on a  setup of interacting spins similar to the proposal of \cite{PhysRevApplied.8.024030}, sufficient conditions for this kind of universality were reported. 
In particular, the convergence property is guaranteed when the norm of the dynamical map is constrained. Recently, a similar definition of convergence was introduced in \cite{higherorderqrc}, where the sufficient condition is provided by the eigenvalues of the dynamical map and a quantum version of the echo state property is introduced.
A different variation of the QRC spin system in terms of quantum circuits, as depicted in Figure~\ref{fig2}(e), was envisioned in \cite{9029180}, being the precedent of the IBM experimental implementation of \cite{PhysRevApplied.14.024065}. In this last work, a
family of QRC circuit models has been identified where
the universality property holds. 
The efforts to analyze and optimize the performance of the original spin model are still ongoing \cite{Kutvonen2020,martinez2020information,martinez2021}.
Regarding QELM systems, the universal approximation property of continuous functions has been studied in~\cite{arXiv:2009.00298v2},  where several scenarios where considered.

All the previously mentioned spin-based references fall into the category of CQC in our classification of Table~\ref{table1}, focusing on solving classical temporal tasks in the QRC framework. Going beyond classical tasks, discrete-variable models of fermions, as depicted in Figure~\ref{fig2}(c), have been introduced to solve non-temporal quantum tasks \cite{universalqrc,9153954,Ghosh2019}. The authors of \cite{Ghosh2019} proposed a 2D Fermi-Hubbard model with nearest-neighbor couplings as a substrate, where each lattice site is pumped by incoherent excitations, while the input is written in two bosonic modes that are  coupled to the reservoir via the dissipative ``cascaded formalism". The whole system is modelled by a Lindblad master equation and the output is built as a function of the fermionic occupation numbers. 
This formalism allows the authors to perform the task of recognizing entangled states and it is potentially useful for any estimation of nonlinear functions of the input state. The same substrate is employed in~\cite{9153954}, harnessing the time multiplexing technique to increase the number of output nodes for quantum state tomography and can be applied to reconstruct both fermionic and bosonic states. Finally,  the Fermi-Hubbard model is also employed in \cite{universalqrc}, where a quantum reservoir model is proposed such that a universal set of quantum gates can be realized. 
We point out variational quantum circuits consisting of coupled superconducting transmons, which are depicted in Figure~\ref{fig2}(d), as another promising candidate for QELM, since fully trained variational quantum circuits solving non-temporal tasks have already been implemented on such platforms \cite{havlivcek2019supervised}.
Elementary building blocks for a quantum counterpart of a network of spiking neurons intended to process quantum input were proposed in \cite{kristensen2019artificial} however, based on simple spin Hamiltonians that implement the transformations under certain conditions involving in particular the interaction time.

\subsubsection{Continuous Variables}

We now move to discuss the substrates that are usually modelled as continuous variables, as for instance optical systems, already well-established versatile and powerful classical RC platforms \cite{brunner2019photonic}.
As for machine learning with quantum continuous variables, proposals to realize either basic building blocks \cite{benatti2019continuous} or full scale quantum NN \cite{killoran2019continuous} using variational (i.e. parameterized) quantum circuits have recently been reported. In these proposals, linear transformations and displacements of the input are achieved by Gaussian gates.
Nonlinearity is introduced with non-Gaussian gates in \cite{killoran2019continuous}. Measurement-induced nonlinearity is proposed as an alternative to experimentally difficult non-Gaussian gates in \cite{benatti2019continuous}, however no advantage over classical perceptrons was observed. Although intended to be trained by adjusting the circuit parameters, these proposals could be adapted to realize continuous-variable QELM or QRC by starting from a randomly initialized but fixed feedforward or recurrent architecture, respectively, and training only the readout layer.
A first proposal in this direction is the use of a Gaussian boson sampler as a QELM \cite{wright2020capacity}, where classical input could be encoded in, e.g., squeezing and quantum input directly injected and the output would be a function of the expected values of detected photons.
Also for ELM, nonlinear waves substrates have been recently shown to display  universal approximator properties under certain conditions  \cite{PhysRevLett.125.093901}, being attainable in nonlinear optical media and Bose-Einstein condensates.

As for QRC, the case of Gaussian states of random linear networks to perform classical tasks, i.e. CQC in Table~\ref{table1}, was proposed in \cite{nokkala2020gaussian}. In this approach, nonlinearity originates from the input and readout layers whereas the network state provides memory. The input layer consists of a generally nonlinear mapping of the classical input sequence into quantum states, which are then injected into the network with periodic state resets of one of the network oscillators. The observables of the rest of the oscillators are combined with a trained function to match a desired output. Already with these minimal resources, an improvement over purely classical resources was found (see Section~\ref{performance} for more details), and the reservoir family was found to have the universal approximation property. Here, in contrast to spin-based models, the linear dynamical equations lead to a necessary and sufficient condition for convergence \cite{nokkala2020gaussian}. This approach is suited for realization in photonic devices, promising candidate platforms to be used as substrates for information processing. 
The different frequency or temporal modes found in photonic systems, as depicted in Figure~\ref{fig2}(f), could serve are reservoir for QRC and QELM. 
For instance, it was recently proposed the generation of re-configurable complex networks in Ref.~\cite{Nokkala_2018njp}
based for instance on frequency combs~\cite{Cai2017} interacting through nonlinear media, as we represent in Figure~\ref{fig2}(f).
As a further example, lattices of coplanar waveguide resonators have already been produced for photons in the microwave regime, as in~\cite{Kollar2019}. This kind of systems are experimentally flexible platforms where different geometric configurations are possible between the network connections, such as hyperbolic lattices. The discussion in~\cite{Kollar2019} also points to the possibility of including interactions between photons by means of nonlinear materials, and of coupling the resonators to superconducting qubits, ideal for QRC and QELM.

A different scheme, based on a single Kerr nonlinear oscillator, for CQC was adopted in \cite{qrcsnonlinearoscill}.
Although only a single observable is considered, the scheme allows for its continuous monitoring;  the output signal was discretized to facilitate training with the usual methods, leading to a situation not unlike the temporal multiplexing mentioned earlier for the spin based approach. First considered for QELM \cite{qrcsnonlinearoscill}, more recently, also temporal tasks have been reported in this scheme \cite{kalfus2021neuromorphic}.
Substrates that can be modelled using continuous variables have also been used to tackle quantum tasks. A random bosonic lattice of nonlinear nodes excited by a randomly distributed classical field has been proposed for a CQQ \cite{PhysRevLett.123.260404}
scheme of QELM.
Following the principles of ELMs and RCs, only the readout layer, i.e. the basis change implemented by the linear optics, is trained.\\

%%%%%%%%%%%%%%%%%%%%%%%%%%%%%%%%%%%%%%%%%%%%%%%%%%%%%%%%%
\subsection{Examples of Classification and Temporal Tasks}
\label{examples}

In Section~\ref{tasks}, we described different computational tasks with a focus on their classical or quantum nature without making an explicit distinction between tasks that belong to QRC from those that are proper to QELM. Based on the differences underlined in Section~\ref{RCandELM}, we can find examples of both approaches. In particular, several works reports on temporal tasks employing the memory of the reservoir, a defining feature of QRC ~\cite{PhysRevApplied.8.024030,PhysRevApplied.11.034021,Kutvonen2020,PhysRevApplied.14.024065,Chen2019,9029180,higherorderqrc,qrsmlnisq,nisqstockforecast,martinez2020information,martinez2021,nokkala2020gaussian,rcqsmeasurement,kalfus2021neuromorphic}. 
In contrast, when the tasks that are performed do not require memory from past inputs to generate the output, we are dealing with a QELM~\cite{universalqrc,9153954,Ghosh2019,PhysRevLett.123.260404,KRISNANDA2021141,negoro2018machine,qrcsnonlinearoscill,kalfus2021neuromorphic}. 
Below, for the sake of concreteness, we present an  example of squeezing classification using QELM and an example of a temporal task realized with QRC.

\subsubsection{QELM Classifier}

\begin{figure}[H]
\includegraphics[width=0.95\columnwidth]{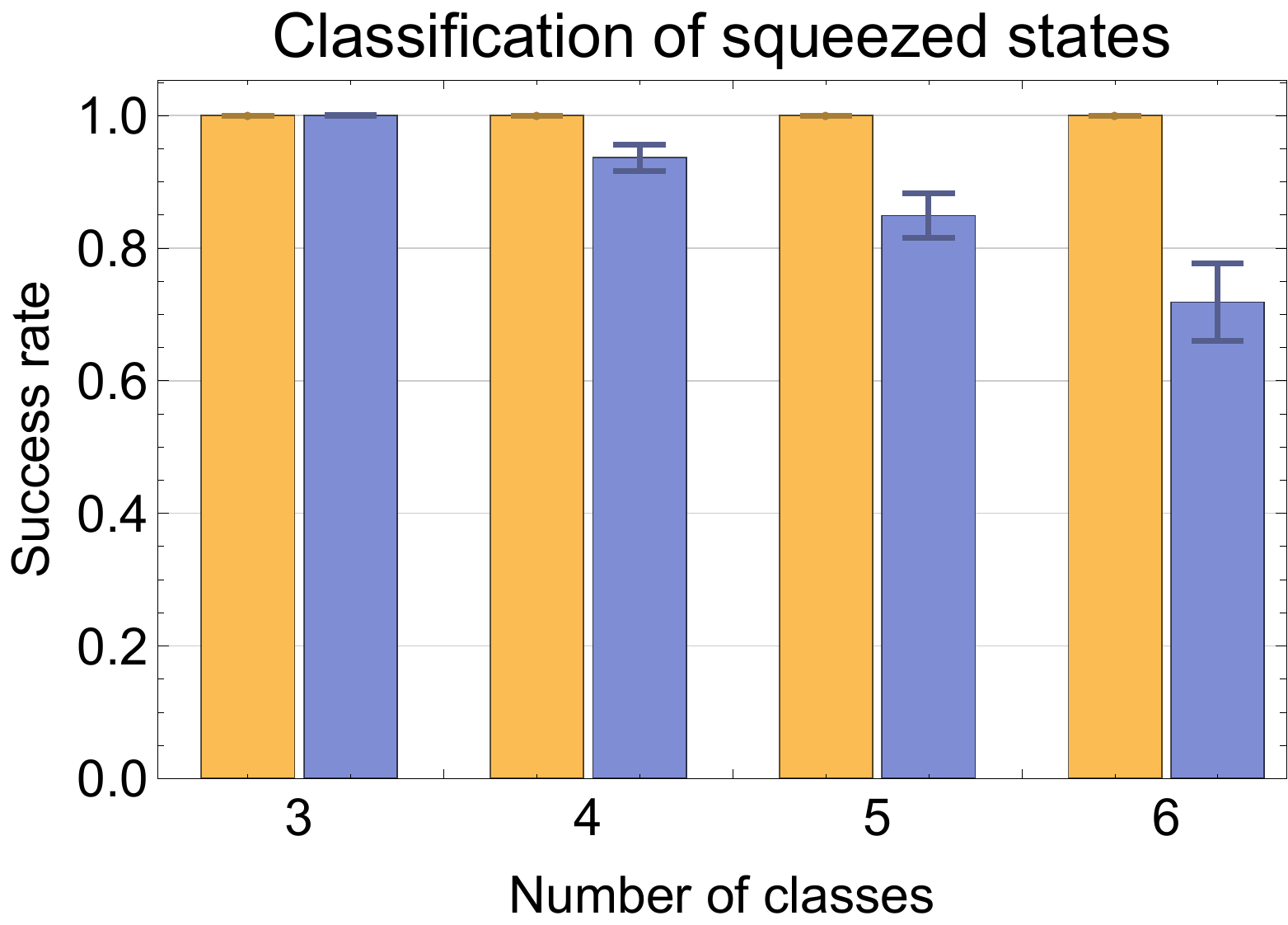}
\caption{Squeezed vacuum states with either a fixed (yellow, left bars) or a random (blue, right bars) phase are classified according to the magnitude of squeezing using a QELM. 100 random realizations of both the states to be classified and of the oscillator network acting as QELM are considered for each case. The columns and error bars indicate the success rate and standard deviation, respectively.}
\label{fig:squeezing}
\end{figure}

Dynamics of quantum systems that do not retain memory of past inputs can be used to solve non-temporal tasks such as classification, where the inputs are categorized following some given rule. In Figure~\ref{fig:squeezing}, we show results of using the harmonic oscillator network of Ref.~\cite{nokkala2020gaussian} in such a fashion to classify squeezed states.

Specifically, the input is a squeezed vacuum state with squeezing magnitude $r\leq 2$, which can take a finite number of different values, and phase $\varphi$, which is either constant or selected uniformly at random from the interval $[0,\pi/4]$. The different classes are simply the different values for $r$, and the oscillator network is trained to classify the states accordingly.
\noindent
The network consists of $N=4$ oscillators, initially in its ground state. The input state is injected into the network by setting the state of one of the oscillators to it. The network is then allowed to evolve for a fixed time before the output is extracted. The output, i.e. the class assigned to the input state, is a trained function of 6 of the network observables. As can be seen, if the phase is constant the network succeeds in the task in all cases. With random phase the success rate remains high for a small number of classes. This simple classification example displays the versatility of QRC platforms also serving for QELM: indeed the difference with the original model studied in Ref.  \cite{nokkala2020gaussian} is merely that here we reset the state of the network to the ground state between inputs and with the purpose of a classification task. Further details are given in Appendix~\ref{app_classifier}.

\subsubsection{QRC  Timer}\label{sec:QRCexample}

The natural dynamics of quantum systems can be employed to solve temporal tasks. Here, we show the performance of the quantum spin model of Ref. \cite{PhysRevApplied.8.024030}  in one of the benchmark tasks used to characterize the memory capacity of reservoir computers: the timer task.

\begin{figure}[H]
\includegraphics[width=0.95\columnwidth]{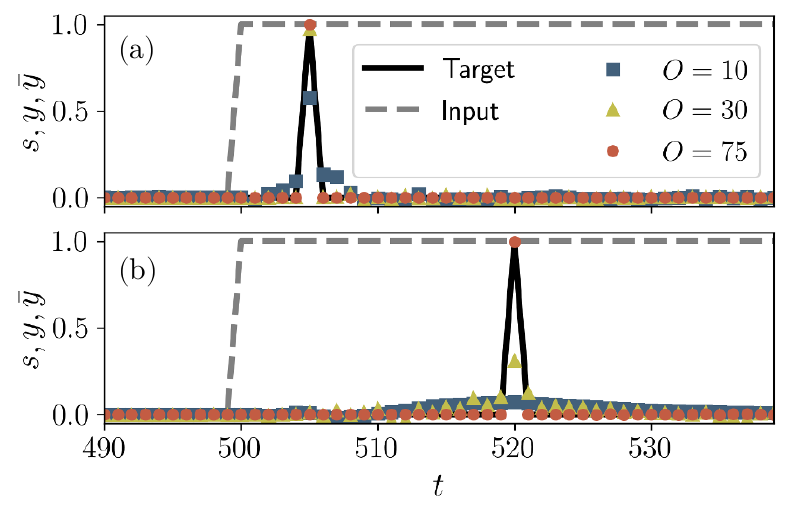}
\caption{Timer task trajectories for a quantum spin network of $N=10$ spins. In panel (a) the countdown time is $\tau=5$, whereas in panel (b) we used $\tau=20$. The input sequence $\{s_k\}$ (dashed line), the target $\{\bar{y}_k\}$ (solid line), and outputs $\{y_k\}$ (symbols) are shown in both situations.
We used three different instances of the output layer with different number of $O$ observables in each instance, namely $O=10$, squares, $O=30$, triangles, and $O=75$ circles (see Appendix~\ref{app_timer} for more details).}\label{timer}
\end{figure}

The timer task is a relatively simple task that consists in obtaining a response from the system after a countdown finishes. The indication to start the countdown is given by the input sequence $\{s_k\}$ (dashed-grey line) going from zero to one, and the countdown finishes when the target sequence $\{\bar{y}_k\}$ shows a spike (black-solid line). Figure~\ref{timer}(a) and (b) account for different countdown times (5 and 20 time steps, respectively), testing the limits of the memory capabilities of our system. Three instances of the output layer were employed for a network of $N=10$ spins, where $O$ denotes the number of observables. These output layers were constructed with different combinations of local
\noindent
observables, see Appendix~\ref{app_timer} for further details. In both situations, $O=75$ outperforms
the smaller output layers, observing a drastic decrease of capabilities for $O=10$ and $O=30$ when we extend the countdown time to $\tau=20$. These results are a simple but convincing evidence of the positive influence on the system performance of using a larger number of observables \cite{martinez2020information}. The possibility to use a large number of observables relies on the large number of exploitable degrees of freedom that quantum systems contain.

%%%%%%%%%%%%%%%%%%%%%%%%%%%%%%%%%%%%%%%%%%%%%%%%%%%%%%%%%
\subsection{Performance of Quantum Substrates }
\label{performance} 
 
A key motivation in considering quantum substrates in machine learning is the possible advantage these can represent with respect to classical ones. In this section, we summarize some recent results on the performance both in QELM and in QRC. Where available, we report on the performance of a quantum substrate and its classical counterpart on the execution of the same task. 
A potential advantage of the extension of RC and ELM to the quantum domain is the possibility of dealing with an exponentially large number of degrees of freedom even for small networks of coupled units. 
The way these degrees of freedom can be computationally exploited will not only depend on the size of the Hilbert space of the reservoir, but also on the way information is encoded in, processed by and extracted from the system.
 
The study of the computational power in relation with the system size is indeed the target of several works on QRC. In the pioneering work of Ref. \cite{PhysRevApplied.8.024030}, a spin network was employed to solve problems such as the short-term memory task and the parity check. These tasks serve to estimate the capacity of the system to store the input sequence and to perform a nonlinear mapping, respectively.
By tuning the input injection rate and the coupling strength, spin networks of a few qubits ($N<10$) exhibit similar computational capabilities to echo state networks (ESN) of $100-500$ nodes.
The performance of a $500$-node ESN can be matched by a spin network of only $N=7$ qubits, for instance measuring at several times along the response of each qubit.
The possibility to extract additional information from the dynamics of each spin at different times emphasizes the role played by the hidden degrees of freedom of the system.  
%IPC example:
In a similar spirit, the potential of exploiting the large amount of computational nodes that are available in a quantum substrate  was quantified in \cite{martinez2020information} through the IPC, a tool well suited to assess the performance in RC (see Section \ref{crcomp}). 
Taking as observables quantum correlations as well as spin projections, and considering time multiplexing techniques, it was demonstrated that the capacity of spin networks can be extended without indications of saturation. Moreover, it has been shown that different sets of observables may provide distinctive (linear and nonlinear) contributions to the capacity~\cite{martinez2020information}.

Indeed, the total computational capability increases with the output size if we employ linearly independent observables~\cite{martinez2020information}. 
In order to fully benefit from a large Hilbert space, all of them must have access to the input either directly or through the interconnections.
Otherwise, saturation can occur. For the single-qudit reservoir system in Ref. \cite{kalfus2021neuromorphic}, although the quantum reservoir outperforms its classical counterpart, the performance quickly saturates as the number of considered levels grows. This indicates that richer dynamics should be considered to exploit the whole Hilbert space~\cite{kalfus2021neuromorphic} and to achieve an improved performance.

The relevance of the state space size and encoding is also manifested in \cite{nokkala2020gaussian}, where a network of quantum harmonic oscillators is compared with an ESN with the same number of computational nodes, i.e., the same number of oscillators as the number of nodes in the ESN. The quantum system possesses a larger state space even when restricted to Gaussian states. Furthermore, the way in which the input is encoded in the quantum network plays a determinant role in exploiting all degrees of freedom. Computing the IPC for systems with $N=8$ computational nodes (both for the ESN and the quantum network), it was shown that encoding in the mean amplitude increases the total capacity only by a factor of 2 respect to the ESN.
In contrast, encoding in the fluctuations of the input oscillator allows to access to $N^2$ observables. In particular, squeezed vacuum encoding increases the nonlinear memory of the system, while classical thermal fluctuations only provide linear memory. 
%%%%%%%%%%%%%%%%%%%%%%%%%%%%%%%%%%%%%%%%%%%%%%%%%%%%%%%%%

As we have seen, in the case of classical tasks and outputs the performance achievable with quantum substrates is quantified using  the same tools that have been  developed for
 classical RC and ELM,  and direct comparisons with classical substrates can be done. On the contrary,  for  quantum tasks we have seen a broader spectrum of approaches and anticipated some of their specific advantages in Section \ref{quantumsubstrates}. 
In the case of the quantum tasks mentioned above (entanglement detection, quantum state tomography, quantum state preparation)
no  RC or ELM protocols have been devised so far that are based on classical substrates (to the best of our knowledge, Ref. \cite{rcqsmeasurement} is the only example of a classical RC approach to a quantum problem). 
In order to assess the usefulness and the advantage of the  protocols proposed in Refs. \cite{Ghosh2019,PhysRevLett.123.260404,KRISNANDA2021141}, a comparison can be made with previous results on the same tasks that have been obtained using more standard (fully-trained) classical NN architectures.
As an example, the task of classifying, through a  NN, whether a (multipartite) state is entangled or not  was  addressed in Ref. \cite{qstaba34cbib22}. There, the classification is limited to pure states and the extension to mixed ones represents a nontrivial issue. In contrast, the simple model of \cite{Ghosh2019}, based on a quantum substrate, works irrespective of the mixedness of the state under study.
As for quantum state preparation, the QELM of \cite{PhysRevLett.123.260404} does not require any additional resource apart from the reservoir itself, in contrast with other NN proposals, which for instance rely on number-resolved measurements \cite{ODriscoll2019}, or
conditional measurements \cite{jpcoab524abib42}. 

The impact of noise has also been addressed in most of the platforms for QRC and QELM. In qubit based reservoirs, both effects such as dephasing \cite{PhysRevApplied.8.024030,Chen2019} and additive external noise in observables \cite{PhysRevApplied.8.024030,higherorderqrc,9153954} have been considered. Ref. \cite{Chen2019} compared effects of dephasing to (generalized) amplitude damping and found them to be similar, whereas Ref. \cite{9153954} compared systematic and random noise, finding that the reservoir learns to compensate for the former and tolerates the latter. Importantly, noise does not necessarily rule out universal QRC \cite{PhysRevApplied.14.024065} and it has been suggested \cite{nisqstockforecast} that a small amount of noise can prevent overfitting in NISQ systems, similarly to the classical neural networks case. Sometimes computation speed can be sacrificed for better noise tolerance by simply increasing the number of repetitions for the protocol: this approach was found to make nontrivial information processing possible even when the signal-to-noise ratio was low in an NMR experiment \cite{negoro2018machine}.  Additive external noise has also been considered in continuous-variable systems \cite{PhysRevLett.123.260404,qrcsnonlinearoscill,killoran2019continuous,IEEEpaper} and it has been found that performance degrades gracefully. In particular, impact of external noise in observables can be reduced if the memory of the reservoir can be tuned to be close to the requirements of the task at hand \cite{IEEEpaper}. Furthermore, simulations suggest that a quantum neural network implemented in an optical platform can learn to compensate for photon loss \cite{killoran2019continuous}. 

\section{Experimental and Theoretical Challenges} 
\label{challenges}

Many opportunities arise when RC and ELM are extended into the quantum regime but still there are several open challenges.
One of the main drawbacks for experimental implementations of QRC and QELM comes from the output extraction. 
In most cases, the readout data taken from the quantum substrates are typically expectation values of local node observables or their correlations. This implies that, due to the stochastic nature of quantum measurements, several copies of the system or multiple measurements are generally needed.
Furthermore, in  temporal tasks, the system output needs to be measured in a sequence of times. As  measurements introduce a (even strong) back-action in the reservoir dynamics, their effect needs to be accounted for in on-line time series processing. In order to avoid it, the protocol would have to be repeated many times until each time step that one wishes to consider.  
Finally, the interaction with external degrees of freedom of the quantum reservoir induces decoherence and dissipation in it. These issues are important when dealing with quantum substrates and become particularly challenging for on-line time series processing as in QRC.

A way of dealing with the measurement problem consists in taking a large number of quantum substrates and performing ensemble computing, as considered in the first experimental implementation of QELM on NMR platforms \cite{negoro2018machine},
where the observable averages are taken over a large number of identical molecules in a solid. Back-action effects are negligible in those cases, as weak measurements are performed over the substrate~\cite{Laflamme}.
Although promising, no implementations of temporal data processing (QRC) have been reported on NMR platforms so far. 

NISQ schemes for QRC are also promising and have also been experimentally implemented on current IBM quantum computers \cite{PhysRevApplied.14.024065,nisqstockforecast}. The output of quantum computation on the IBM platform is so far obtained through projective measurements, so the measurement back-action problem discussed above represents a serious issue when compared with on-line processing in classical RC. For instance, the experimental implementation reported in Ref. \cite{PhysRevApplied.14.024065} required for each time a different ensemble of realizations, a rather laborious procedure; the system needs be restarted from time $0$ to be  measured at time $ k \Delta t$  so the substrate retains memory from the past.

Most theoretical proposals stand on the assumption that it is possible to average over several identical realizations. The number of repetitions required  to obtain a reliable estimation of expectation values will depend on the system as well as on the observable considered.  For instance, the estimation of variances in continuous variables of quantum vacuum states will require, in general, more effort than the estimation of a mean quadrature of an intense mode with small quantum fluctuations \cite{qrcsnonlinearoscill}.
Depending on the specific features of each platform, different strategies for output extraction and measurement can be devised and this represents an almost completely unexplored field. Alternative implementations in which statistics can be performed over a single realization would be a significant advance in the field. In the case of QRC, experimental platform proposals allowing on-line data processing are also desirable. These frameworks could be based on sequential measurements of the substrate, implemented in a way in which the reservoir does not completely lose all information about past inputs after each detection. This would most likely yield back-action effects on the dynamics of the substrate, which should also be taken into account.

A field so far unexplored in QRC is represented  by the processing of temporal sequences of inputs that can bring information, for instance, about the stochastic source that generates them.
For this class of problems, the measurement back-action problem is in general less  worrying than in the  time series processing protocols described so far. Indeed, if  information were encoded in  the entire time series, the final state of the substrate would be determined by the whole input injection process, so that an output could be built  at the final computing stage  after a single measurement. This class of problems is not necessarily limited to processing of classical inputs, as it can also include quantum time series processing. For instance, one may want to determine whether a quantum channel is Markovian or not, or whether noise in a multipartite quantum register is correlated or uncorrelated. This family of cases would represent QRC tasks within the QQQ cell of Table \ref{table1}. 

Broadly speaking, a general theory for quantum tasks, $T_{Q}$, is yet to be formulated. In this regard, a new concept analog to the universal approximation property should be defined. For instance, in the case of quantum state tomography \cite{9153954}, a universality property would ensure that any arbitrary quantum state belonging to a given  Hilbert space can be reconstructed. 
Furthermore, there is still no theoretical magnitude that can estimate the performance of a given quantum substrate by computing how well it approximates any arbitrary solution, similar to the IPC in classical RC that provides an overall assessment  of the linear and nonlinear memory properties of the system \cite{dambre2012information}. A step towards this direction has been recently reported in \cite{wright2020capacity} where an information-theoretic quantifier of the memory of a quantum or classical learning machine processing quantum or classical inputs is introduced.
Although its use for such tasks is discussed, it has yet to be applied in this manner.

Finally, in situations where the output is a quantum state, training cannot be done with linear regression because the readout layer might be, e.g., an interaction Hamiltonian \cite{universalqrc}, which also makes such class of protocols highly unorthodox with respect to the standard definition of reservoir computing.
There might also be additional restrictions for the trained parameters to be admissible.
Although in such situations general purpose optimization methods, such as the Nelder-Mead algorithm, have been successfully used to train the system~\cite{universalqrc}, it could be wondered if theoretical work on the optimal training method could further improve results in these cases.

%%%%%%%%%%%%%%%%%%%%%%%%%%%%%%%%%%%%%%%%%%%%%%%%%%%%%%
\section{Conclusions \& Outlook}
\label{conclusions}

This review article on QRC and QELM covers the state-of-the-art and provides an outlook of future research avenues in this emerging field.
We have introduced a classification of the current literature in terms of the classical or quantum nature of the input, substrate and task, respectively, that provides a general overview and points out unexplored directions.
Recent works have shown that several classical and quantum tasks can be successfully performed.
A relevant feature of QRC and QELM is the combination of the benefits of simple training requirements, compared to other machine learning methods, with the potential of improved performance of quantum substrates.
Several platforms are good candidates to be used as quantum substrates and facilitate the practical implementation in currently available NISQ devices. An additional envisioned advantage of the QRC framework applied to NISQ devices is to bypass the strict requirements of quantum error correction or error mitigation techniques, since time-invariant readout errors can be learned during training and are compensated for by the output layer \cite{PhysRevApplied.14.024065}.
Further opportunities remain to be explored  including photonic platforms, particularly successful for classical RC  \cite{brunner2019photonic}.

We note that quantum substrates allow one to tackle problems that are incompatible with their classical counterparts.
Nowadays, conventional classical machine-learning algorithms are being used for tasks related to assisting quantum experimental realizations. Tasks include e.g. quantum state preparation, quantum state tomography, calculation of entanglement-related quantities from accessible observables, and continuous monitoring of systems.
QRC and QELM proposals widen the opportunities as can be more naturally adapted to interact with quantum inputs or embedded in experimental platforms.
According to the present state of the art, while QRC has been mainly employed in classical temporal tasks, QELM has found its realm in quantum information tasks. However, the use of QRC is expected to display a disruptive potential in quantum temporal tasks as, for instance, in quantum secure communication, quantum channel tomography, non-Markovianity detection, or quantum error correction. In this line, a very recent proposal for temporal quantum process tomography is found in~\cite{arXiv:2103.13973v2}.
Classical or quantum RC substrates embedded in quantum systems can indeed allow for low-latency processing of quantum signals at the computational edge with high fidelity \cite{rcqsmeasurement}, operating at the timescales of the measured quantum system and providing a speed-up.

To conclude, the field of QRC and QELM is still moving its first steps, so it is premature to make quantitative comparisons with their  much  more advanced classical counterparts in terms of efficiency or performance, especially because of the absence of fully working experimental implementations. However, the perspective we have presented here gives us a glimpse of the potential and the versatility of these frameworks together with the new avenues that deserve to be explored.

%%%%%%%%%%%%%%%%%%%%%%%%%%%%%%%%%%%%%%%%%%%%%%%%%%%%%%%%%
\section*{Acknowledgements}
We acknowledge the Spanish State Research Agency, through the Severo Ochoa and Mar\'ia de Maeztu Program for Centers and Units of Excellence in R\&D (MDM-2017-0711) and through the  QUARESC project (PID2019-109094GB-C21 and -C22/ AEI / 10.13039/501100011033);
 We also acknowledge funding by CAIB through the QUAREC project (PRD2018/47).
The work of MCS has been supported by MICINN/AEI/FEDER and the University of the Balearic Islands through a ``Ramon y Cajal'' Fellowship (RYC-2015-18140). GLG is funded by the Spanish  Ministerio de Educaci\'on y Formaci\'on Profesional / Ministerio de Universidades   and  co-funded by the University of the Balearic Islands through the Beatriz Galindo program  (BG20/00085).

%%%%%%%%%%%%%%%%%%%%%%%%%%%%%%%%%%%%%%%%%%%%%%%%%%%%%%%%%
\section*{Conflict of Interest}
The authors declare no conflict of interest.
%%%%%%%%%%%%%%%%%%%%%%%%%%%%%%%%%%%%%%%%%%%%%%%%%%%%%%%%%
\appendix
\section{Appendix}
\subsection{Details on the QELM Classifier}
\label{app_classifier}

The classification task of Figure~\ref{fig:squeezing} is defined by
\begin{equation}
    \begin{split}
       \rho_{in}&=\ket{r,\varphi}\bra{r,\varphi},\\
        \bar{y}&=r,
    \end{split}
\end{equation}
where the input state, $\rho_{in}$, is a single-mode Gaussian state in squeezed vacuum, determined by squeezing magnitude $r\in[0,2]$ and phase $\varphi\in[0,\pi/4]$. Here $r$ can take only a finite number of equally spaced different values which are the classes. For three classes the possible values are $r\in\{0,1,2\}$ and so on. The phase $\varphi$ is either uniformly random or takes a constant value $\varphi=0$. The target output $\bar{y}$ is the class, and should coincide with $r$.

The starting point of the used QELM is the oscillator network presented in Section~IIA of \cite{nokkala2020gaussian}, where it was used for QRC. Here it is converted into a QELM by resetting its state to the ground state between inputs, removing any memory of input history. The network is completely connected and consists of $N=4$ oscillators. The bare frequencies of all oscillators is set to $\omega_0=0.25$, whereas the interaction strengths in the network are chosen uniformly at random from $g\in[0,0.2]$. At the beginning of the protocol the state of the oscillator selected for input injection is set to $\rho_{in}$. The network is allowed to evolve for a time $\Delta t$, after which the diagonal elements of the covariance matrix of the rest of the oscillators are used to form the output. The value of $\Delta t$ is chosen by considering a large set of possible values and choosing the one that minimizes the spectral radius specified in Lemma 1 of \cite{nokkala2020gaussian}. The network is first trained as in \cite{nokkala2020gaussian} (except that there is no preparation phase since the network state is reset between inputs) using random input states. Typically the outputs are real numbers close to $r$. The predicted class is taken to be the nearest class to the network output; the training is finished by shifting the bias term such that this leads to the correct classification as often as possible. The trained weights are then tested by classifying a new set of random inputs.

The success rate in the test phase is shown in Figure~\ref{fig:squeezing}. The network is first trained with $500$ inputs and then used to classify $200$ fresh inputs. Both inputs with a constant phase $\varphi=0$ and with a random phase are considered and the number of classes, i.e. possible values for $r$, is varied. $100$ random realizations of both the inputs and the networks were considered for each of the $8$ different cases.

%%%%%%%%%%%%%%%%%%%%%%%%%%%%%%%%%%%%%%%%%%%%%%%%%%
\subsection{Details on the QRC Timer}
\label{app_timer}
The input/output map of the timer task of Figure~\ref{timer} can be represented by
\begin{equation}
\begin{split}
&s_k=\begin{cases}
  1 &   k \geq c,  \\
  0 & k<c,
 \end{cases} \\
 &\bar{y}_k=\begin{cases}
  1 &   k = c +\tau,\\
  0 & \text{otherwise},
 \end{cases}
\end{split}
\end{equation}
where the values of the inputs $\{s_k\}$ indicate when the countdown starts ($k=c=500$). Once the countdown is initialized, the target is to obtain an isolated response of the output sequence with value $\bar{y}_{c+\tau}=1$. 

The system we considered here is the quantum spin model of \cite{PhysRevApplied.8.024030}. We used a network of $N=10$ spins, with an homogeneous external magnetic field $h=10$, coupling strength between qubits given by a random uniform distribution $J_{ij}\in [-1/2,1/2]$ and input injection rate $\Delta t=10$, everything in normalized units. Three different instances of the output layer were employed in Figures~\ref{timer}(a) and (b), where $O$ denotes the number of observables. First, blue squares correspond to the case of using only the spin projections in the $z$-axis $\braket{\sigma^z_i}$ ($1\leq i\leq N$). Second, yellow triangles correspond to all the local spin projections $\braket{\sigma^x_i}$, $\braket{\sigma^y_i}$ and $\braket{\sigma^z_i}$. Third, the red dots were obtained with the two-spin correlations along the $z$-axis $\braket{\sigma^z_i\sigma^z_j}$  $(1\leq i,j\leq N$, $i< j)$ in addition to all the local spin projections.

Figures~\ref{timer}(a) and (b) represent the results for the timer tasks of $\tau=5$ and $\tau=20$, respectively. We fed an input sequence of 800 time steps to the system, where the first 400 were a used as warming up to remove the initial condition. The output layers were trained only with the last 400 time steps, and we took 10 different realizations of the couplings of the network to obtain averages of the output trajectories $\{y_k\}$.

%%%%%%%%%%%%%%%%%%%%%%%%%%%%%%%%%%%%%%%%%%%%%%%%%%
%
\subsection{Details on the IPC}
\label{app_IPC}

Dambre et al. recently introduced the 
information processing capacity (IPC) \cite{dambre2012information} as a tool to estimate
the computational capabilities of a classical dynamical system.
They demonstrated that the total computational capacity of a dynamical system is bounded by the number of linearly independent variables that are used for the output, and this bound is saturated when the dynamical system has fading memory, i.e. when the dynamical system dissipates the information about the input after some time.

To estimate the computational capabilities of a system, the capacity to approximate products of orthogonal functions, like the Legendre polynomials that can depend on present and past inputs, is evaluated. More specifically, the target function for a given degree $d$ of nonlinearity is:
\begin{equation}\label{Eq:Pol}
    \bar{y}_k=\prod_i \mathcal{P}_{d_i}[s_{k-i}], \quad \sum_{i}d_i=d,
\end{equation} 
such that the sum of the degrees $d_i$ of all the multiplied polynomials add up to a given degree $d$ and
where the inputs $s_{k-i}$ are randomly and uniformly distributed.

Then, the capacity of approximating target functions like Eq.~\ref{Eq:Pol} is quantified as
\begin{equation}\label{Eq:C} C_L(X,\textbf{y})=1-\frac{\text{min}_{\textbf{w}}\text{MSE}_L(\textbf{y},\bar{\textbf{y}})}{\braket{\bar{\textbf{y}}^2}_L},
\end{equation}
where $X$ is the matrix of the dynamical variables at different times, $\textbf{y}$ and $\bar{\textbf{y}}$ are the prediction and target sequences respectively and $\textbf{w}$ is the vector of weights of the output layer. The mean square error MSE is the cost function $\text{MSE}_L(\textbf{y},\bar{\textbf{y}})=\frac{1}{L}\sum^L_{k=1}(y_k-\bar{y}_k)^2$ and the bracket $\braket{}_L$ denotes the temporal average for sequences of length $L$.

The total capacity is exactly saturated when infinite input sequences are considered and the contributions of all the possible nonlinearities, i.e. up to an infinite degree $d$, are computed. However, a sufficiently long input sequence and sufficiently high maximum degree give a stable result.

A first application of the IPC to QRC has been reported for a spin reservoir in Ref. \cite{martinez2020information}.

%BIBLIOGRAPHY

%%%%%%%%%%%%%%%%%%%%%%%%%%%%%%%%%%%%%%%%%%%%%%%%%%%%%%%%%
\end{multicols}
%%%%%%%%%%%%%%%%%%%%%%%%%%%%%%%%%%%%%%%%%%%%%%%%%%%%%%%%%
\end{document}